\documentclass[dvips,12pt]{article}
\usepackage{times,latexsym,amssymb}
\usepackage[dvips]{graphicx,color}

\newcommand{\ignore}[1]{}

\newcommand{\ket}[1]{|{#1}\rangle}
\newcommand{\bra}[1]{\langle{#1}|}
\newcommand{\kets}[2]{|{#1}\rangle_{{}_{\!{#2}}}}
\newcommand{\bras}[2]{{}_{{}_{{#2}\!}}\langle{#1}|}
\newcommand{\slb}[2]{{{#1}^{({#2})}}}

\newcommand{\cE}{{\cal E}}

\newcommand{\cG}{{\cal G}}

\newcommand{\cL}{{\cal L}}

\newcommand{\cM}{{\cal M}}

\newcommand{\cS}{{\cal S}}

\newcommand{\tensor}{\otimes}
\newcommand{\trace}{\mbox{tr}}

\newtheorem{Theorem}{Theorem}

\newtheorem{Corollary}[Theorem]{Corollary}
\newcommand{\proof}{\paragraph*{Proof.}}

\newcommand{\qed}{\hspace*{\fill}\rule{2.5mm}{2.5mm}%
\vspace*{8pt}\par}

\newcounter{boxfigure}

\newcounter{natboxfig}

\newcounter{natfig}

\newlength{\isplength}
\newcounter{ispcount}

\begin{document}

\title{Fermionic Linear Optics and Matchgates\\
\normalsize Extended Abstract}

\author{E. Knill
\\
{\tt knill@lanl.gov}}
\date{}
\maketitle

\begin{abstract}
Fermionic linear optics is efficiently classically simulatable.  Here
it is shown that the set of states achievable with fermionic linear
optics and particle measurements is the closure of a low dimensional
Lie group. The weakness of fermionic linear optics and measurements
can therefore be explained and contrasted with the strength of bosonic
linear optics with particle measurements.  An analysis of fermionic
linear optics is used to show that the two-qubit matchgates and the
simulatable matchcircuits introduced by Valiant generate a monoid of
extended fermionic linear optics operators. A useful
interpretation of efficient classical simulations such as this one is
as a simulation of a model of non-deterministic quantum computation.
Problem areas for future investigations are suggested.
\end{abstract}

\section{Introduction}

It is conjectured that standard quantum computation is more efficient
than probabilistic computation. The conjecture is supported by the
ability to efficiently factor large numbers~\cite{shor:qc1995a} and
simulate physics~\cite{lloyd:qc1996a} using quantum computers, by
proofs that quantum computers are more powerful with respect to some
black boxes~\cite{simon:qc1997a}, and by results showing exponential
improvements in communication complexity~\cite{raz:qc1999a}.

To delineate the conjecture one can consider models of computation
where the basic operations are multiplication of linear operators in
a given set $G$. Each operator in $G$ is associated with a complexity
(e.g. the length of its name), so that the complexity of a product
$g_1g_2\ldots$ is the sum of the complexities of the $g_i$.  One can
then ask questions about the complexity of calculating quantities like:
1. Computing the entries in a standard basis of a
product. 2. Computing the trace of a product. When $G$ is the a set of
elementary quantum gates, the power of quantum computers is equivalent
to being able to efficiently sample from a probability distribution
with expectation an entry of a product and variance $O(1)$
(see~\cite{knill:qc1998c}). The power of one-bit quantum
computers~\cite{knill:qc1998c} is equivalent to sampling from a
probability distribution with expectation the trace of a product and
variance $O(2^n)$, where $n$ is the number of qubits.

A special case is when the set $G$ is the group of operators
normalizing the group generated by the Pauli matrices (bit flip, sign
flip). For $n$ qubits, this group has order $2^{O(n^2)}$ and plays a
crucial role in encoding and decoding stabilizer
codes~\cite{cleve:qc1996b} and in fault tolerant quantum
computation~\cite{gottesman:qc1997a}.  In~\cite{gottesman:qc1998a} it
is shown that even when this group is extended by projections onto the
logical states of qubits, the complexities of the two questions above
are polynomial. Two similarly defined groups consist of the linear optics
operators for fermions and for bosons.  In both cases, the groups are Lie
groups of polynomial dimension in the number of modes. (Modes play the
same role as qubits in these systems).  A few simulatability results
were known for these groups.  For example, for bosons, the orbit of
the vacuum state under the linear optics operators consists of Gaussian
states, for which many relevant quantities can be efficiently
computed.  Similarly, particle preserving linear optics operators applied to
exactly one boson lead only to states that are equivalent to classical
waves~\cite{reck:qc1994a,wallentowitz:qc2000a}.

Recently, Valiant~\cite{valiant:qc2001a} demonstrated a set of
products of operators (those definable by a class of
``matchcircuits'') for which the complexities of the first question
and many of its generalizations are polynomial. Terhal and
DiVincenzo~\cite{terhal:qc2001a} realized that this set includes the
unitary linear optics operators for fermions and that as a consequence, it
is unlikely that it is possible to realize quantum computation in
fermions by means of linear optics operators and particle detectors with
feedback. They give a direct and efficient simulation of these
operators based on fermionic principles.  This result is at first
surprising: In~\cite{knill:qc2000e} it was shown that with bosons,
linear optics operators and particle detectors with feedback are sufficient
for realizing quantum computation.  The difference between fermions
and bosons is explained by realizing that the effects of particle
detectors are expressible as limits of non-unitary linear optics operators
in fermions but not in bosons. As a result, the states achievable with
fermionic linear optics operators and particle measurements are in the
closure of a ``simple'' set.

Since matchgate operators are non-unitary, one can ask what
additional power is provided by Valiant's simulation of
matchgates. Here it is shown that the two-qubit matchgates densely
generate the monoid given by the closure of a group of extended
fermionic linear optics operators in the Jordan-Wigner
representation~\cite{jordan:qc1928a}.  This group defines the
non-deterministic computations that can be physically realized with
unitary linear optics operators and particle measurements.  The equivalence
of two-qubit matchgates and fermionic linear optics two-qubit operators
generalizes to the set of simulatable matchcircuits introduced by
Valiant.

\section{Fermionic Linear Optics}

Let $I,\slb{X}{k},\slb{Y}{k},\slb{Z}{k}$ denote the identity and the
Pauli operators acting on qubit $k$.  Define $U_k =
\slb{Z}{1}\ldots\slb{Z}{k-1}\slb{U}{k}$ ($U_1=\slb{U}{1}$) for
$U=X,Y$. Then the $U_k$ define a representation of fermionic mode
operators. In particular, $(X_k+iY_k)/2$ and $(X_k-iY_k)/2$ represent
the annihilation and creation operators for mode $k$. Let $\cL_1$ be
the linear span of the identity together with the $U_k$ for $1\leq
k\leq n$, where $n$ is the number of modes (or qubits).  The set $\cG_1$
of \emph{fermionic linear optics operators} is the set of invertible
matrices that preserve $\cL_1$ by conjugation.  That is, $g\in\cG_1$
iff for all $A\in\cL_1$, $gAg^{-1}\in\cL_1$. The terminology refers to
the property that conjugation of an annihilation or a creation
operator results in a linear combination of such operators.  Let
$\cL_2$ be the set of products of two operators in $\cL_1$, so that
$\cL_2 = \cL_1\cL_1$.  The group $\cG_2$ of \emph{extended} linear
optics operators is the set of invertible matrices that preserve
$\cL_2$. Note that $\cG_1\subseteq\cG_2$.  (In bosons, the analogous
definitions lead to identical groups.)  The group $\cG_2$ is
considered to be ``unphysical'' for fermions, due to the presence of
odd products of annihilation and creation operators. In
Sect.~\ref{sect:g1=g2} it is shown that $\cG_2$ is naturally viewed as
a subgroup of $\cG_1$ for one more mode.

The space $\cL_2$ is a (complex) Lie algebra. It is spanned by the Pauli
operator products given by $I$,
$U_k$, $\slb{Z}{k}$ , and
$\slb{U}{k}\slb{Z}{k+1}\ldots\slb{Z}{k+l}\slb{V}{k+l+1}$ with
$U,V\in\{X,Y\}$. The dimension of
$\cL_2$ is $2n^2+n+1$. By considering general sums of Pauli products,
one can check that if for every $A\in\cL_2$, $[X,A]\in\cL_2$, then
$X\in\cL_2$. It follows that $\cL_2$ is the Lie algebra of $\cG_2$.
All strictly quadratic (in $\cL_1$) terms of $\cL_2$, together with
the identity also form a Lie algebra $\cL_2'$ of dimension $2n^2-n$,
which is the Lie algebra of $\cG_1$.  Physically, realizable
operators are continuously generated from the identity. As a result,
for the remainder of the paper, $\cG_i$ is assumed to be given by the
exponentials of $\cL_i$.\footnote{Without a proof that this assumption
holds, it is possible that the groups studied here are only the
component of the identity of the originally defined groups.}

In using (extended) linear optics operators for computation, one starts with
the vacuum state $\ket{\mathbf{v}_n}=\kets{0\ldots0}{1\ldots n}$ and
applies operators in $\cG_1$ ($\cG_2$) and measurements in the number
basis $\kets{0}{k},\kets{1}{k}$ for a mode. The outcomes of
measurements are given by applying the measurement projections
$\kets{0}{k}\bras{0}{k}={1\over 2}(I+\slb{Z}{k})$ and
$\kets{1}{k}\bras{1}{k}={1\over 2}(I-\slb{Z}{k})$. For standard
computation, which projection ``happens'' is determined by the square
amplitude of the result of applying it. For non-deterministic
computation we can ``choose'' the outcome.  In either case, analysis
of the capabilities requires studying products of operators in $\cG_i$
and the measurement projections.  Let $\cS_i$ be the monoid given by
the topological closure of $\cG_i$.

If $\cG_2$ and measurements could be used for efficient faithful
quantum computation, then the set of states $S_n$ obtained with such
operators from the $n$-mode vacuum state has to contain sufficiently
large subspaces. That is, the $2^m$ dimensional state space of $m$
qubits must be contained in $S_n$ with $n=O(\mathrm{poly}(m))$.  The
following theorem makes this unlikely.

\begin{Theorem}
\label{thm:cs2clos}
The monoid generated by measurement projections
and $\cG_2$ is contained in $\cS_2$.
\end{Theorem}

\proof
This is a consequence of the fact that the measurement
projections are limits of elements of $\cG_2$:
\begin{eqnarray}
{1\over 2}(I+\slb{Z}{k})
            &=& \lim_{t\rightarrow\infty} e^{t\slb{Z}{k}}/e^t\nonumber\\
{1\over 2}(I-\slb{Z}{k})
            &=& \lim_{t\rightarrow\infty} e^{-t\slb{Z}{k}}/e^{t}
  \label{eq:lim_ident}
\end{eqnarray}
\qed

Since $\cG_2$ is a $2n^2+n+1$-dimensional Lie group,
Thm.~\ref{thm:cs2clos} implies that $\cS_2\ket{\mathbf{v}_n}$ is the
closure of a small dimensional space. This suggests that $\cS_2$ is
not sufficiently strong for quantum computation.  The fact that the
normalizer of the Pauli group together with standard measurements are
insufficient~\cite{gottesman:qc1998a} follows in a similar way. That
is, applying normalizer operators and projections onto stabilizer
codes to the standard initial state always results in stabilizer
states.

Note that a similar result cannot be shown for bosonic linear
optics operators with particle measurements. Only the projection operator
onto the $0$ boson state of a mode is expressible as a limit of
(non-unitary) linear optics operators.  This provides an explanation of why
efficient linear optics quantum computation is
possible~\cite{knill:qc2000e}.

\section{Matchgates and Linear Optics Operators}

In~\cite{valiant:qc2001a}, Valiant introduced a family of linear
operators (called matchgates) acting on qubits. Matchgates are based
on a graph theoretic construction. Valiant showed that under certain
conditions, the coefficients of matrices defined by products of
matchgates could be efficiently calculated. Matchgates acting on two
qubits were shown to satisfy a set of $5$ equations, the matchgate
identities. If $B$ is the matrix defined by a matchgate acting on two
qubits, then the following are $0$:
\begin{eqnarray}
M_1 &=& \bra{00}B\ket{00}\bra{11}B\ket{11}
-\bra{10}B\ket{10}\bra{01}B\ket{01}
-\bra{00}B\ket{11}\bra{11}B\ket{00}
+\bra{10}B\ket{01}\bra{01}B\ket{10} \nonumber\\ \label{eq:m1}
\\
M_2 &=& \bra{10}B\ket{00}\bra{11}B\ket{11}
-\bra{10}B\ket{10}\bra{11}B\ket{01}
-\bra{11}B\ket{00}\bra{10}B\ket{11}
+\bra{10}B\ket{01}\bra{11}B\ket{10}\nonumber \\ \label{eq:m2}
\\
M_3 &=& \bra{01}B\ket{00}\bra{11}B\ket{11}
+\bra{01}B\ket{01}\bra{11}B\ket{10}
-\bra{11}B\ket{00}\bra{01}B\ket{11}
-\bra{01}B\ket{10}\bra{11}B\ket{01}\nonumber\\ \label{eq:m3}
\\
M_4 &=& \bra{00}B\ket{01}\bra{11}B\ket{11}
+\bra{01}B\ket{01}\bra{10}B\ket{11}
-\bra{00}B\ket{11}\bra{11}B\ket{01}
-\bra{10}B\ket{01}\bra{01}B\ket{11}\nonumber\\ \label{eq:m4}
\\
M_5 &=& \bra{00}B\ket{10}\bra{11}B\ket{11}
-\bra{10}B\ket{10}\bra{01}B\ket{11}
-\bra{00}B\ket{11}\bra{11}B\ket{10}
+\bra{01}B\ket{10}\bra{10}B\ket{11} \nonumber\\ \label{eq:m5}
\end{eqnarray}
Let $\cM_2$ be the set of matrices $B$ satisfying the identities
$M_i=0$ and either $\bra{11}B\ket{11} \not= 0$ or $B$ is diagonal.
Valiant showed that these matrices are realizable
by matchgates.

\begin{Theorem}
\label{thm:cmcs2}
The closure of $\cM_2$ is $\cS_2$ for two modes (or qubits).
\end{Theorem}

\proof
The Lie algebra which densely generates $\cS_2$ is spanned by the
$11$ operators
\begin{equation}
L=\{II,XI,YI,ZI,ZX,ZY,XX,XY,YX,YY,IZ\}
\end{equation}
Here $UV$ abbreviates $\slb{U}{1}\slb{V}{2}$.  One can check that for
$A\in L\setminus\{II\}$, $A (YX) + (YX) A^T = 0$: It suffices to note
that if $A^T=A$, then $A$ anticommutes with $YX$, and if $A^T=-A$,
which is the case if $A$ contains an odd number of $Y$'s, then $A$
commutes with $YX$. (This property generalizes for arbitrary number of
qubits, using the operator $YXYX\ldots$ instead of $YX$.)  The
identity $A (YX) + (YX) A^T = 0$ can be re-written in the form
$(A\tensor I + I\tensor A) T = 0$, where $T$ is the antisymmetric
vector
\begin{equation}
T = \ket{00}\ket{11}-
    \ket{11}\ket{00}+
    \ket{01}\ket{10}-
    \ket{10}\ket{01}.
\end{equation}
This means that $T$ is an eigenvector of the Lie group $\cL$ generated
by $L\oplus L = \{A\tensor I + I\tensor A : A\in L\}$.
Note that $\cL=\{B\tensor B:B\in\cG_2\}$.
$\cL$ preserves antisymmetric vectors, so the statement
that $\cL T\propto T$ is equivalent to 
$R^T\cL T = 0$ for all $R$ antisymmetric such that $R^T T = 0$.
The dimension of such $R$ is $5$, and here is a basis:
\begin{eqnarray}
R_1 &=& \ket{00}\ket{11}-\ket{11}\ket{00}
       -\ket{01}\ket{10}+\ket{10}\ket{01} \\
R_2 &=& \ket{00}\ket{01}-\ket{01}\ket{00} \\
R_3 &=& \ket{00}\ket{10}-\ket{10}\ket{00} \\
R_4 &=& \ket{01}\ket{11}-\ket{11}\ket{01} \\
R_5 &=& \ket{10}\ket{11}-\ket{11}\ket{10}
\end{eqnarray}
Define the expressions
\begin{eqnarray}
E_i &=& R_i^T B T\\
E^T_i &=& T^T B R_i
\end{eqnarray}
Since for two qubits $\cL_2^T = \cL_2$, members $B$ of $\cG_2$ satisfy
the identities $E_i=0,E^T_i=0$.
Because these identities are all
derived from an eigenvector condition, the set of matrices $B$
satisfying them is a closed monoid $\cG'_2$ containing $\cG_2$.

Using the equivalence
\begin{equation}
(\ket{ab}\ket{cd})^T B\tensor B (\ket{ef}\ket{gh})
 = \bra{ a b}B\ket{ e f}\bra{ c d}B\ket{ g h},
\end{equation}
one can check that the following hold
\begin{eqnarray}
E_1+E^T_1 &=& 4M_1\\
E_4 &=& 2M_3\\
E_5 &=& 2M_2\\
E^T_4 &=& 2M_4\\
E^T_5 &=& 2M_5\\
\bra{11}B\ket{11}(E_1-E^T_1) &=& 4(\bra{01}B\ket{11} M_2
        -\bra{10}B\ket{11} M_3
        +\bra{11}B\ket{10} M_4
        -\bra{11}B\ket{01} M_5)\nonumber\\ \\
\bra{11}B\ket{11}E_2 &=& 2(\bra{01}B\ket{11} M_1
        -\bra{00}B\ket{11} M_3
        +\bra{01}B\ket{10} M_4
        -\bra{01}B\ket{01} M_5)\nonumber\\ \\
\bra{11}B\ket{11}E_3 &=& 2(\bra{10}B\ket{11} M_1
        -\bra{00}B\ket{11} M_2
        -\bra{10}B\ket{01} M_5
        +\bra{10}B\ket{10} M_4)\nonumber\\ \\
\bra{11}B\ket{11}E^T_2 &=& 2(\bra{11}B\ket{01} M_1
        -\bra{11}B\ket{00} M_4
        +\bra{10}B\ket{01} M_3
        -\bra{01}B\ket{01} M_2)\nonumber\\ \\
\bra{11}B\ket{11}E^T_3 &=& 2(\bra{11}B\ket{10} M_1
        -\bra{11}B\ket{00} M_5
        -\bra{01}B\ket{10} M_2
        +\bra{10}B\ket{10} M_3).\nonumber\\
\end{eqnarray}
Mathematica instructions to check the above relationships
are included verbatim in Appendix~\ref{app:math1}.

Since diagonal matrices trivially satisfy $E_i=0$, $E^T_i=0$ ($i>1$)
and $E_1-E^T_1=0$, the identities imply that $\cM_2\subseteq\cG'_2$.  Let
$\cM_2'=\{B\in\cM_2:\bra{11}B\ket{11}\not=0\}$.  By directly solving
for the entries of $B$ other than $\bra{11}B\ket{11}$ in the first
summand of the $M_i$, one can see that $\cM_2'$ is an analytically
coordinatizable 11 complex dimensional manifold.  The diagonal members
of $\cM_2$ are in the closure of $\cM_2'$.

The identities also imply that the elements of $\cG_2$, and therefore
those of $\cS_2$, satisfy $M_i=0$. It follows that
the $B\in\cS_2$ with $B$ diagonal or $\bra{11}B\ket{11}\not=0$
are in $\cM_2$.

For invertible $B$, the identities $E_i=0$ imply that $B(XY)B^T =
\lambda XY$ for $\lambda\not=0$. It follows that the tangent space at
$B$ is exactly that of $\cG_2$ at $B$. Consequently, $\cM'_2$ and
$\cG_2$ contain the same invertible matrices satisfying
$\bra{11}B\ket{11}\not=0$. It remains to show that these matrices are
dense in both sets. For $\cM'_2$ it suffices to observe that for fixed
$\bra{11}B\ket{11}\not=0$, there is an invertible $B\in\cM'_2$, which
implies that the determinant function is not null on this linearly
defined subset. Hence the complement of the determinant's null set is
dense.  For $\cG_2$ the density property follows from the fact that
the subgroup generated by $XI$ and $XX$ acts transitively on the basis
states.

\qed

\section{Simulatable Matchcircuits}

Valiant showed that any composition of operators consisting of two
qubit matchgates on the first two qubits and gates of the
form $e^{t(\slb{X}{k}\slb{X}{k+1})}$ and $e^{t(\slb{Y}{k}\slb{Y}{k+1})}$
is efficiently simulatable in the following sense: If $B$ is a product
of $m$ such gates, then many sums of squares or square norms of
entries of $B$ can be computed efficiently in $m$ and $n$ (the number
of qubits). Let $\cM$ be the set of all products of the gates mentioned.

\begin{Theorem}
\label{thm:cmcsn}
The closure of $\cM$ is $\cS_2$.
\end{Theorem}

\proof By definition and by Thm.~\ref{thm:cmcs2},
$\overline{\cM}\subseteq\cS_2$.  It suffices to show that the
invertible operators in $\cM$ generate $\cG_2$. This can be checked
directly by using the Bloch sphere rules for conjugating products of
Pauli matrices by $90^\circ$ rotations ($e^{-iU\pi/4}$) around other
products~\cite{sorensen:qc1983a}.  For example,
$\slb{Z}{1}\slb{Z}{2}\slb{X}{3}$ is obtained by conjugating
$\slb{Z}{1}\slb{Y}{2}$ with a rotation around
$\slb{X}{2}\slb{X}{3}$. The operator $\slb{Z}{3}$ is obtained by
conjugating $\slb{Z}{1}\slb{Z}{2}\slb{X}{3}$ with a rotation around
$\slb{Z}{1}\slb{Z}{3}\slb{Y}{3}$. The latter operator can be deduced
similarly to the way $\slb{Z}{1}\slb{Z}{2}\slb{X}{3}$ was obtained.
Induction can be used to extend to arbitrarily many qubits.  \qed

\section{Non-deterministic Computations}

A non-deterministic computation with fermionic linear optics consists
of a sequence of linear operators and measurements, where one
post-conditions on the measurement outcome in the sense that one
multiplies the state by the appropriate projection operator. The
outcome is not normalized. Let $U$ be the implemented operator. The
minimal quantities one wishes to compute efficiently are
$\trace(\bra{\mathbf{v}_n}U^\dagger(I\pm
\slb{Z}{k})U\ket{\mathbf{v}_n})$, which give the relative
probabilities of the outcome of a measurement on the $k$'th
mode. Suppose that implementable operators form a monoid and include
the standard measurement projections. Since
$\ket{\mathbf{v}_n}\bra{\mathbf{v}_n}=\prod_k((I+Z)/2)$ and
$(I+Z)=(I+Z)^\dagger(I+Z)$, it suffices to be able to compute, for
each implementable $U$, $\trace(U^\dagger U)=\sum_{kl}|U_{kl}|^2$.
This motivates a definition that works for any monoid generated by
\emph{elementary} operators: An \emph{efficient simulation} is defined
to be an efficient algorithm for computing $\trace(U^\dagger U)$ for
an explicitly implemented (as a product of elementary operators)
$U$. Efficiency is defined in terms of the implementation complexity
of $U$.  With this definition, Valiant demonstrated an efficient
simulation of matchcircuits composed of certain matchgates. The
purpose of this section is to discuss how that leads to an efficient
simulation of a dense subset of the monoid $\cS_2$ with naturally
defined generators.

An \emph{elementary fermionic gate} is an operator of the form $\alpha
e^{itU}= q+rU$ with $U$ one of the products of Pauli operators in
$\cL_2$ other than the identity. The coefficients $q$ and $r$ are
required to be complex rationals with $q\not=0$.  Let $d$ be the
number of digits needed to denote these rationals.  The description
length of $q+rU$ is $\Omega(2\log(n)+d)$, where the summand $2n$ is
the description length of $U$, one of $O(n^2)$ many possible Pauli
products. The elementary projection is the operator
$(I+\slb{Z}{n+1})/2$. It is implementable non-deterministically by
post-selection on a particle measurement.

The elementary fermionic gates can be realized in terms of the
operators allowed in simulatable matchcircuits: Simply conjugate
one of these operators by the appropriate sequence of $90^\circ$
allowed operators. Note that the $90^\circ$ operators are elementary
if scaled by $\sqrt{2}$. The standard measurement projections
are allowed in matchcircuits. 
It is therefore possible to take a product of
elementary fermionic gates and projections, and
efficiently express them using allowed matchgates. It follows that
Valiant's algorithm can be used to efficiently simulate the monoid
$\cE_2$ generated by elementary fermionic gates and 
projections.  The goal is to show that these operators densely
generate $\cS_2$.

\begin{Theorem}
\label{thm:u+m}
Except for a scale factor, the operators of $\cG_2$ on $n$ modes
are implementable by first adjoining a mode
in state $\kets{0}{n+1}$, applying
a sequence of unitary operators of $\cG_2$ for $n+1$ modes
and elementary projections and finally discarding mode $n+1$.
\end{Theorem}

\proof It suffices to show that $e^{t\slb{Z}{n}}$ with real $t$ is
implementable up to a scale.  This follows from the observation that
other real exponentials of Pauli operators are conjugates of
$e^{t\slb{Z}{n}}$ by unitary operators, and these together with
unitary operators generate $\cG_2$.

To implement $e^{\pm t\slb{Z}{n}}$ realize the following sequence of operators:
\begin{itemize}
\item[1.] Adjoin $\kets{0}{n+1}$ (if that hasn't already been done).
\item[2.] Apply $e^{is (\slb{X}{n}\slb{X}{n+1}+\slb{Y}{n}\slb{Y}{n+1})/2}$
\item[3.] Project mode $n+1$ with $(I+\slb{Z}{n+1})/2$, which returns mode
$n+1$ to its initial state, or results in $0$.
\end{itemize}
To see how this works, apply it to $\alpha\kets{0}{n}+\beta\kets{1}{n}$.
Step 2 is a partial swap with a phase and results
in $\alpha\kets{0}{n}+\beta(\cos(s)\kets{1}{n}\kets{0}{n+1} +
                            i\sin(s)\kets{0}{n}\kets{1}{n+1})$.
The elementary projection results in 
$(\alpha\kets{0}{n}+\cos(s)\beta\kets{1}{n})\kets{0}{n+1}$.
It follows that the effect is the same as applying
a scalar multiple of $e^{-\ln\left(\cos(s)\right)\slb{Z}{n}/2}$.
The other sign in the exponent can be obtained by
replacing step 2. with:
\begin{itemize}
 \item[2'.] Apply $e^{is (\slb{X}{n}\slb{X}{n+1}-\slb{Y}{n}\slb{Y}{n+1})/2}$
\end{itemize}
\qed

\begin{Corollary}
The closure of $\cE_2$ is $\cS_2\tensor(I+\slb{Z}{n+1})/2$.
\end{Corollary}

\proof This follows from Thm.~\ref{thm:u+m} and the fact that the
elementary rotations $e^{itU}$ for $U$ a Pauli product densely
generate all such rotations.  (See, for
example,~\cite{bernstein:qc1997a}.)
\qed

It can be seen that the ability to efficiently simulate
non-deterministic computation as defined above leads
to an efficient simulation of a quantum computation with
measurements and future operators conditioned on the measurement
outcomes. The method is described in~\cite{terhal:qc2001a}
and basically consists of simulating, at each step,
the random measurement outcome, using a calculation of the conditional
probability distribution.

A potentially easier problem then efficient simulation of a monoid is
to determine, for an implemented $U$, whether $U=0$.
Observe that if it was possible to
use $\cS_2$ with elementary generators to efficiently and faithfully
realize quantum computation, then the zero-test algorithm can be used
to efficiently solve problems in polynomial quantum non-deterministic
time as defined in~\cite{adleman:qc1997a}. In~\cite{fenner:qc1999a} it
was shown that this is hard for the polynomial hierarchy.

\section{Identifying the Lie Algebras: $\cG_1$ is General}
\label{sect:g1=g2}

Let $\cL^{\circ}_2$ be the set of trace zero members of $\cL_2$.  The
adjoint action of $\cG_2$ on $\cL^{\circ}_2$ permits representing
members of $\cG_2$ as $(2n^2+n)\times (2n^2+n)$ matrices. The
representation is faithful up to scalar multiples, because
$\cL^{\circ}_2$ algebraically generates all operators on the $n$
qubits. This means that products of elementary operators can be
efficiently computed in the representation. The reverse procedure,
i.e.  finding a decomposition of a represented operator in terms of a
product of exponentials of Pauli products is also possible, though
less obviously so.  For this purpose it is more useful to recognize
$\cL^{\circ}_2$ as the Lie algebra $\mathfrak{so}_{2n+1}\mathbb{C}$
and work in the fundamental representation. One way to recognize
$\cL^{\circ}_2$ is to realize that it is (isomorphic to ) a
subalgebra of $\cL_2'$ for one more qubit.  The mapping is
accomplished by modifying the members of the form $U_k$ by multiplying
with $\slb{X}{0}$. This makes the operators strictly quadratic for
fermionic modes $0,\ldots,n$ (in this order). Then observe that
$\cL_2'$'s adjoint action on $\cL_1$ is the fundamental representation of
$\mathfrak{so}_{2n+2}\mathbb{C}$. The algebra can now be identified.
Incidentally, this construction shows that in a sense $\cL_2$ is no
more general then $\cL_2'$ despite appearances. This together with the
results of the previous section implies that the simulation algorithm
of Terhal and DiVincenzo~\cite{terhal:qc2001a} can be used to simulate
$\cS_2$ with the same generality as Valiant's.

Here is the direct way to identify $\cL^{\circ}_2$ as a Lie algebra:
In the fundamental representation $\mathfrak{so}_{2n+1}\mathbb{C}$ is
spanned by the antisymmetric matrices $s_{ij}=
\ket{i}\bra{j}-\ket{j}\bra{i}$ for $0\leq i<j\leq n$.  The
identification is made via the correspondences
\begin{eqnarray}
iX_k/2 &\rightarrow& s_{0k}\\
iY_k/2 &\rightarrow& s_{0(n+k)}\\
i\slb{Z}{k}/2 &\rightarrow& s_{k(n+k)}\\
i\slb{X}{l}\slb{Z}{l+1}\ldots \slb{X}{k}/2 &\rightarrow& s_{(n+l)k}\\
i\slb{X}{l}\slb{Z}{l+1}\ldots \slb{Y}{k}/2 &\rightarrow& s_{(n+l)(n+k)}\\
i\slb{Y}{l}\slb{Z}{l+1}\ldots \slb{X}{k}/2 &\rightarrow& -s_{lk}\\
i\slb{Y}{l}\slb{Z}{l+1}\ldots \slb{Y}{k}/2 &\rightarrow& -s_{l(n+k)}
\end{eqnarray}

This identification of $\cL^{\circ}_2$ permits efficiently
representing a product of elementary operators as a
$(2n+1)\times(2n+1)$ matrix. Let $A$ be a matrix thus obtained. Then
$A^TA=I$, and this identity characterizes the Lie group generated by
the $s_{kl}$. The process of representing a matrix satisfying $A^TA=I$
as a product of elementary operators is straightforward by using a
variant of Gaussian elimination to represent $A$ as a product of
$O(n^2)$ matrices of the form $e^{i t s_{kl}} =
(I+s_{kl}^2)-\cos(t)s_{kl}^2+i\sin(t)s_{kl}$ ($t$ may be complex). By
using conjugation rules by $90^\circ$ rotations, one can then 
expand this into a $O(n^3)$ product consisting only of operators that
are allowed for Valiant's simulatable matchcircuits.

\section{Concluding Comments}

It is true that bosons can be represented by paired fermions. So why
does this not lead to an efficient realization of quantum computers by
using this representation together with techniques for bosonic linear
optics?  One answer is that the bosonic linear optics operators in this
representation correspond to Hamiltonians that are quartic in the
annihilation and creation operators and are therefore not in $\cL_2$.
It is in fact not hard to see that adding to $\cL_2$ only the
Hamiltonian $\slb{Z}{1}\slb{Z}{2}$, the Lie algebra generated contains
all products of Pauli matrices and so generates all invertible
matrices~\cite{bravyi:qc2000a}.

\vspace*{\baselineskip}

\noindent{\bf Suggested problem areas for future investigations:} 
\begin{itemize}
\item[1.] Determine the complexity of efficiently simulating
representations of the three families of simple complex Lie groups.
Is the complexity polynomial in the dimension of the groups?
  
Notes:
  \begin{itemize}
  \item[] The results of Valiant, Terhal and Divincenzo and this paper
  show that the answer is ``yes'' for  one family of representations.
  \item[] The answer might depend on the choice of generators and
  elementary operators. The fundamental representation of each
  such group can be used to make a reasonably natural definition.
  \item[] Which projectors in a representations are to be assumed
  as elementary operators? They should be in the closure of the Lie
  group.
  \item[] Semisimple Lie algebras can be analyzed in terms of their simple
  parts. What about non-semisimple ones?
  \end{itemize}

\item[2.] What finite monoids of operators are efficiently simulatable?

Notes:
  \begin{itemize}
   \item[] Again, the choice of generators may be crucial, and it is
   desirable that it is ``natural'' in some sense.
   \item[] The monoids associated with $n$-ary stabilizer codes
   via the appropriate normalizer are efficiently simulatable
   in terms of the number of systems used.
   \item[] Is the stabilizer code example naturally generalizable?
   \end{itemize}

\item[3.] Problem areas 1. and 2., but for efficiently determining
whether a product of generators is zero. Is this sometimes
strictly easier to do?

\item[4.] Find a group or monoid of operators where the probabilistic behavior
of a (quantum) computation is efficiently simulatable, but
the non-deterministic behavior is not.

Notes:
  \begin{itemize}
  \item[] It is necessary to define what is meant
  by ``probabilistic'' behavior. The one case where an interpretation is
  readily available is if the group is unitary and the initial state 
  as well as standard measurements are provided.
  For a monoid, one approach is to allow as measurements some or all
  partitions of unity definable by its operators. The monoid
  should be (densely) generated by its unitary operators and 
  projections associated with measurements.
  \end{itemize}

\end{itemize}

\appendix
\section{Checking the Matchgate Identities}
\label{app:math1}

\begin{verbatim}
(* Mathematica notes. *)

(* Useful rules: *)
Unprotect[Dot];
Dot[tensor[a_,b_],tensor[c_,d_]] = (a.c)*(b.d);
Dot[-a_,b_] = -(a.b);
Dot[a_,-b_] = -(a.b);
Dot[-a_,-b_] = (a.b);

(* For obtaining the equation for the transpose: *)
trnsprls = {b[c_].k[d_] -> b[d].k[c]};
(* For obtaining the equation for the conjugate by XX: *)
xxrls = {x00->x11,x01->x10,x10->x01, x11->x00};
(* Swapping: *)
swprls = {x01->x10,x10->x01};
lswprls = {b[x01]->b[x10],b[x10]->b[x01]};

(* Conventions:
*  b[xab] stands for $\bra{ab}$, k[xab] for $\ket{ab}$.
*  Quadradic expressions for a matrix B are expressed
*  $\trace(X (B\tensor B))$ with X in the appropriate
*  tensor product space. X is given for various expressions.
*  This way the expression (b[x00].k[x00])*(b[x11].k[x01])
*  refers to the product $\bra{00}B\ket{00}\bra{11}B\ket{01}$.
*)

(* Matchgate expressions: *)
M1 = tensor[b[x00],b[x11]].tensor[k[x00],k[x11]] +
     - tensor[b[x10],b[x01]].tensor[k[x10],k[x01]] +
     - tensor[b[x00],b[x11]].tensor[k[x11],k[x00]] +
     + tensor[b[x10],b[x01]].tensor[k[x01],k[x10]];
M2 = tensor[b[x10],b[x11]].tensor[k[x00],k[x11]] +
     - tensor[b[x10],b[x11]].tensor[k[x10],k[x01]] +
     - tensor[b[x11],b[x10]].tensor[k[x00],k[x11]] +
     + tensor[b[x10],b[x11]].tensor[k[x01],k[x10]];
M3 = tensor[b[x01],b[x11]].tensor[k[x00],k[x11]] +
     + tensor[b[x01],b[x11]].tensor[k[x01],k[x10]] +
     - tensor[b[x11],b[x01]].tensor[k[x00],k[x11]] +
     - tensor[b[x01],b[x11]].tensor[k[x10],k[x01]];
M4 = tensor[b[x00],b[x11]].tensor[k[x01],k[x11]] +
     + tensor[b[x01],b[x10]].tensor[k[x01],k[x11]] +
     - tensor[b[x00],b[x11]].tensor[k[x11],k[x01]] +
     - tensor[b[x10],b[x01]].tensor[k[x01],k[x11]];
M5 = tensor[b[x00],b[x11]].tensor[k[x10],k[x11]] +
     - tensor[b[x10],b[x01]].tensor[k[x10],k[x11]] +
     - tensor[b[x00],b[x11]].tensor[k[x11],k[x10]] +
     + tensor[b[x01],b[x10]].tensor[k[x10],k[x11]];

(* Check: 
M3 - (M4/.trnsprls)
 * = 0 *
 *
M2 - (M5/.trnsprls)
 * = 0 *
*)

(* Lie expressions: *)
T = tensor[k[x00],k[x11]] - tensor[k[x11],k[x00]] +
    tensor[k[x01],k[x10]] - tensor[k[x10],k[x01]];
R1 = tensor[b[x00],b[x11]] - tensor[b[x11],b[x00]] +
     tensor[b[x10],b[x01]] - tensor[b[x01],b[x10]];
R2 = tensor[b[x00],b[x01]] - tensor[b[x01],b[x00]];
R3 = tensor[b[x00],b[x10]] - tensor[b[x10],b[x00]];
R4 = tensor[b[x01],b[x11]] - tensor[b[x11],b[x01]];
R5 = tensor[b[x10],b[x11]] - tensor[b[x11],b[x10]];

E1 = Distribute[R1.T];
ET1 = E1/.trnsprls;
E2 = Distribute[R2.T];
ET2 = E2/.trnsprls;
E3 = Distribute[R3.T];
ET3 = E3/.trnsprls;
E4 = Distribute[R4.T];
ET4 = E4/.trnsprls;
E5 = Distribute[R5.T];
ET5 = E5/.trnsprls;
(* Check:
Simplify[E1+ET1  - 4*M1]
 * = 0 *
 *
Simplify[E4 - 2*M3]
 * = 0 *
 *
Simplify[E5 - 2*M2]
 * = 0 *
 *
Simplify[ (b[x11].k[x11])* E2 -
  2* (
    (b[x01].k[x11])*M1 +
   -(b[x00].k[x11])*M3 +
    (b[x01].k[x10])*M4 +
   -(b[x01].k[x01])*M5
  ) ]
 * = 0 *
 *
Simplify[ (b[x11].k[x11])* E3 -
  2* (
    -(b[x00].k[x11])*M2 +
    -(b[x10].k[x01])*M5 +
     (b[x10].k[x10])*M4 +
     (b[x10].k[x11])*M1
  ) ]
 * = 0 *
 *
Simplify[(M1/.trnsprls) - M1]
 * = 0 *
 *
Simplify[(M2/.lswprls) - M3]
 * = 0*
 *
Simplify[(E2/.lswprls)-E3]
 * = 0*
 *
Simplify[(b[x11].k[x11])*(E1 - (E1/.trnsprls)) -
  4* (
     b[x01].k[x11]*M2 +
    -b[x10].k[x11]*M3 +
    -b[x11].k[x01]*M5 +
     b[x11].k[x10]*M4
  ) ]
 * = 0*
 *
 * This confirms the identites claimed in the text.
*)

\end{verbatim}

\end{document}